\documentclass[preprint]{acm_proc_article-sp}
\usepackage{epsfig}
\usepackage[square,comma,sort&compress]{natbib}

\newcommand{\BigOh}{\mathcal{O}}
\newcommand{\Root}{\mathtt{root}}
\newcommand{\length}{\mathtt{length}}

\begin{document}
\title{Insertion Sort with \\Self-reproducing Comparator P System}
\numberofauthors{3}
\author{
\alignauthor Davood P.Y.B.\\
   \affaddr{LPU--Laguna}\\
   \affaddr{Calamba City, Laguna}\\
   \email{davoodpb@gmail.com}
\alignauthor N.P. Balba\\
   \affaddr{LPU--Laguna}\\
   \affaddr{Calamba City, Laguna}\\
   \email{neilbalba@yahoo.com}
\alignauthor J.P. Pabico\\
   \affaddr{Institute of Computer Science}\\
   \affaddr{UP Los Ba\~nos}\\
   \email{jppabico@up,.edu.ph}
}
\date{}
\toappearbox{15th National Conference on Information Technology Education\\Leyte Normal University, Tacloban City, Leyte\\19--21 October 2017}
\maketitle

\begin{abstract}
We present in this paper a self-reproducing comparator P~system that simulates insertion sort. The comparator~$\Pi_c$ is a degree--2 membrane and structured as $\mu = [_{h_0} [_{h_1}]_{h_1} [_{h_2}]_{h_2} ]_{h_0}$. A maximizing~$\Pi_c$ compares two multisets~$a$ and ~$b$ where $\min(|a|,|b|)$ is stored in compartment~$h_1$ while $\max(|a|,|b|)$ is stored in compartment~$h_2$. A conditional reproduction rule triggers~$\Pi_c$ to clone itself out via compartment division followed by endocytosis of the cloned compartment. We present the process of sorting as a collection of transactions implemented in hierarchical levels where each level has different concurrent or serialized steps.
\end{abstract}

\section{Introduction}
Membrane computing~(MC) is a computer science theoretical discipline which aims to develop new computational models from the physiological processes in biological cells, particularly cellular membranes. The acceptance of MC among theoretical computer scientists grows swiftly starting in 1998 when Gheorghe Paun introduced the idea that a biological entity possesses computing capability~\citep{01-paun06}. The initial goal of MC was to learn from the seemingly computational aspects of the physiological processes in biological cells. 

Paun officially proposed MC in 2000~\citep{02-paun00} and after that various types of membrane systems -- known as P~systems -- were defined, all of them inspired from the computational aspects of bio-chemical processes. The theoretical aspects translated to practical applications as numerous researchers report its applicability to solving computational problems in biomedicine, linguistics, computer graphics, economics, approximate optimization, and cryptography. The understanding of MC has hasten significantly with the introduction of several software products for simulating and implementing P~systems such as SNUPS (simulator of numerical P~systems) and SRPSUMGPU (simulation of recognizer P~systems by using Manycore GPU)\citep{03-gutierrez06,04-martinez09}.

\subsection{Insertion Sort}

Insertion sort is a comparison-based algorithm in which the elements of the input list are sorted one at a time. In this algorithm, the sorted sub-list is always maintained in the lower position of the list. Then the new item is inserted into the previous sorted sub-list such that the new sub-list is also sorted.

We consider the list with one item as a sorted list. Then we iterate, considering one item of the list each repetition and growing the sorted sub-list. In other words, at each iteration, we work on an item from the input list and find its position in sorted sub-list comparing and swapping (if needed) the item with the elements of sorted sub-list started from the end. The iteration stops when swapping stops. The new sub-list is sorted and ready for the next iteration and new item from the list~\citep{09-bender06,10-knuth98,11-astrachan03,12-bentley99}.

Insertion sort is slow compared to the advanced sorting algorithms such as quicksort, merge sort and heapsort~\citep{09-bender06}. Based on time complexity and number of comparison, insertion sort is as slow as bubble sort is; the worst case and the average case for both of insertion sort and bubble sort is~$\BigOh(n^2)$ while the best case is $\BigOh(n)$ where~$n$ is the number of elements of the input list. Although regarding time complexity, insertion sort is weak as bubble sort is, it has some strengths that make it bright despite being disregarded by some computer scientists~\citep{10-knuth98,11-astrachan03}. The strengths of insertion sort can be listed as:

\begin{enumerate}
\item {\bf Simplicity}: Jon Bentley used C programming language and implemented insertion sort only in three lines. He also implemented the optimized version of insertion sort in five lines~\citep{12-bentley99}.
\item {\bf Adaptiveness}: insertion sort efficient for lists in which the elements are already considerably sorted. In this case the time complexity of insertion sort is~$\BigOh(nm)$ when each element in the input list is no more than $m$~places far from its sorted position~\citep{09-bender06,10-knuth98,11-astrachan03,12-bentley99}.
\item {\bf Stability}: insertion sort does not change the relative order of items in the input list with equal indices~\citep{09-bender06,10-knuth98,11-astrachan03,12-bentley99}.
\item {\bf In-placement}: insertion sort needs only one extra unit memory space for swapping of two elements of the input list. The extra unit is small as size of elements of input list.
\item {\bf Online-ness}: insertion sort sorts a list as it receives the elements of the list one-by-one~\citep{10-knuth98,11-astrachan03,12-bentley99}.
\end{enumerate}

\section{Cell-like P~Systems}

\begin{figure}
\centering\epsfig{file=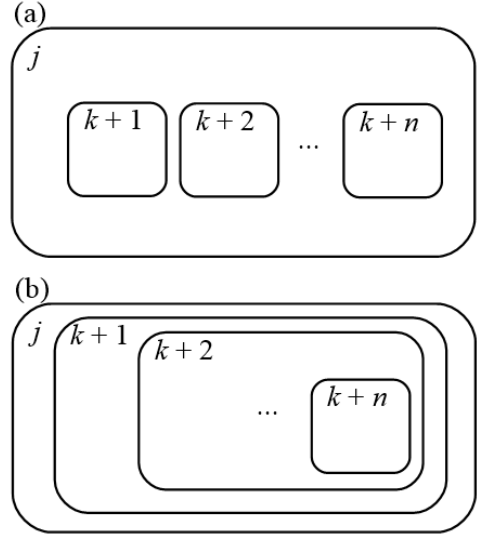, width=2.5in}
\caption{Two general types of hierarchical structure of membranes rooted at $[_i ]_i$: (a)~flat structure; and (b)~deep structure.}\label{fig:membrane-structure}
\end{figure}

Cell-like P~system\footnote{P~System throughout this text for brevity} consists of many membranes arranged hierarchically. These membranes bound compartments. The compartments are the area where multisets of abstract objects are placed. The multisets are sets of objects (or symbols) with multiplicities, while the objects are the ``chemicals'' in the compartments, ``swimming'' in some substance in liquid form~\citep{05-ardelean08, 06-wu16}. The compartments are identified with its index~$i$ and is symbolized as $[_i ]_i$. A membrane~$j$ with~$n$ compartments inside it and structured as a flat rooted tree at~$j$ (Figure~\ref{fig:membrane-structure}a) can be written as 
\begin{equation}
[_j [_{k+1} ]_{k+1} [_{k+2} ]_{k+2} \cdots [_{k+n} ]_{k+n} ]_j.
\end{equation}
Meanwhile, a deep rooted membrane~$i$ with one compartment inside it but that in itself deep-rooted (Figure~\ref{fig:membrane-structure}b) is symbolized as
\begin{equation}
[_i [_{j+1} [_{j+2} \cdots [_{j+n} ]_{j+n} \cdots ]_{j+2} ]_{j+1} ]_i.
\end{equation}
A membrane with only one compartment inside it is both flat- and deep-rooted. In general, a membrane's structure is a combination of these two basis structures.

A multiset can be seen as a string~$S$ where its multiplicity (number of each symbol $|S|=\length(S)$) is significant, not the order of symbols $s\in S$. Since the objects are swimming and moving freely inside a compartment, the permutation or order of objects is not important~\citep{06-wu16,07-singh14,08-chen15}. For example, consider the multisets $M_1 = abbbac=$, $M_2=bacabb$, and $M_3=cbbaab$. All of these multisets are the same (i.e., $M_1=M_2=M_3$) because in each of the multisets, the number of object~$a$, shortened as $|a|$ is two, $|b|=3$, and $|c|=1$. As it can be observed the order of objects in multisets are not important since the objects are inside the liquid and can freely move.

It is noticeable that, in each compartment there are some rules~$R$ and the objects inside the compartment evolve according to these rules. The number of objects in the multiset may change based on the application of the rules. Moreso, the rules do not only direct how the objects change but also how the objects communicate across membranes~\citep{06-wu16,07-singh14,08-chen15}. Although, Paun and others proposed some desirable rules based on the behavior of biological cells, all of the rules presented in the literature so far pay focus on the objects inside the membranes, while the membrane itself is left behind. However, there may be situations in which a cell may reproduce another cell or a membrane may reproduce another membrane. Such situations may find some computational meaning to MC. This is why we propose in this effort a rule for self-replicating membranes. 

In this study, aside from the existing rules that (1)~change the number or type of objects in the membrane and (2)~change the locations of the objects with respect to a membrane, we propose a third rule that allows a membrane to reproduce. In the following subsections, these three types of rules are explained in details.

\subsection{Transmuting Objects}

The first type of rules change some number of objects to some another number of the same or different objects. We call this change a ``transmutation.'' For example, consider the rule $R_1: ca \rightarrow d$ and the objects $aabbbc$ in our previous multisets $M_1$, $M_2$ and $M_3$ that we introduced above (i.e., all multisets have the string $aabbbc$). With the application of $R_1$, this string will be changed to another multiset $M_4=abbbd$. According to $R_1$, one~$a$ and one~$c$ are {\em transmuted} to one~$d$. It is clearly seen that based on $R_1$, the objects and their number from the original multiset were changed: $|a|=1$, $|b|=3$ and $|d|=1$ (we do not talk about object~$c$ any more, since $|c|=0$). In general, given two strings $S_n=s_1s_2\dots s_n$ and $S_m=s_{n+1}s_{n+2}\dots s_{n+m}$, a transmuting rule has the form 
\begin{equation}
s_1s_2\dots s_n \rightarrow s_{n+1}s_{n+2}\dots s_{n+m}\label{eqn:rule-1} 
\end{equation}
where the objects a $s_1, s_2, \dots, s_n$ in an $n$--long string $S_n$ were transmuted into objects $s_{n+1}, s_{n+2}, \dots, s_{n+m}$ of a different $m$--long string $S_m$.

\subsection{Translocating Objects}

Some rules change the locations of the objects. These rules transfer the objects from some membranes to some other membranes in a manner similar to how two or more processes exchange data in a method called {\em interprocess communication}. In physiological processes, common to all biological systems, such transfer is called as {\em cytosis}. In this paper, we call this change as ``translocation,'' which is illustrated by the following example. 

Consider two rules $R_2: a \rightarrow e_{\mathrm [in]}$ and $R_3: d \rightarrow f_{\mathrm [out]}g$ that are applied independently to our example multiset $M_4$ above with string $abbbd$ that is located in membrane $[_{\mathrm focus} ]_{\mathrm focus}$. The membrane is located in a deeply-rooted membrane whose structure is defined by $[_{\mathrm out} [_{\mathrm focus} [_{\mathrm in} ]_{\mathrm in} ]_{\mathrm focus} ]_{\mathrm out}$. The multiset is transmuted to $bbbg$, since object~$a$ is in itself transmuted to~$e$ and at the same time translocated to another compartment $[_{\mathrm in} ]_{\mathrm in}$ inside the current membrane $[_{\mathrm focus} ]_{\mathrm focus}$. Meanwhile, object~$d$ is transmuted to objects~$f$ and~$g$, with~$f$ transferred outside to $[_{\mathrm out} ]_{\mathrm out}$ of the current membrane $[_{\mathrm focus} ]_{\mathrm focus}$. In general, a translocating rule has the form 
\begin{equation}
s_1s_2\dots s_n \rightarrow s_{n+1[i+1]}s_{n+2[i+2]}\dots s_{n+m[n+m]}.\label{eqn:rule-2} 
\end{equation}

In biological processes, a translocation of objects from $[_{\mathrm focus} ]_{\mathrm focus}$ to $[_{\mathrm in} ]_{\mathrm in}$ is called as {\em endocytosis}, while from $[_{\mathrm focus} ]_{\mathrm focus}$ to $[_{\mathrm out} ]_{\mathrm out}$ is called {\em exocytosis}.

\subsection{Cloning Membranes}

At some situations that could be useful in computation, we allow a membrane~$\mu_1$ to reproduce by cloning itself to another membrane~$\mu_2$. Cloning allows the duplication of~$\mu_1$ to several copies of itself, similar to how the biological cells divide in a process called {\em mitosis}. Three scenarios can be inferred from the cloning action with respect to the initial location of the cloned membrane: (1)~outside, (2)~beside, and (3)~inside of the cloned membrane. Let $R_{\mathrm out}$, $R_{\mathrm side}$, and $R_{\mathrm in}$ be the rules that respectively define these scenarios. Further, let $\mu_1 = [_1 \mu_z ]_1$ be the original membrane to be cloned, $\mu_2 = [_2 ]_2$ be the cloned membrane, and $\mu_z$ represent one or more membranes within $\mu_1$. Then, in general:

\begin{eqnarray}
R_{\mathrm out}  &:& [_1 \mu_z ]_1 \rightarrow [_2 [_1 \mu_z  ]_1 \mu_z ]_2\label{rule:clone-out}\\
R_{\mathrm side} &:& [_1 \mu_z ]_1 \rightarrow [_1 \mu_z ]_1 [_2 \mu_z ]_2\label{rule:clone-side}\\
R_{\mathrm in}   &:& [_1 \mu_z ]_1 \rightarrow [_1 [_2 \mu_z ]_2 \mu_z ]_1\label{rule:clone-in}
\end{eqnarray}

\subsection{Computation by Rule Application}

The rules belonging to all three types can be implemented and applied in many ways. These rules emulate biological processes wherein biochemical reactions happen in concurrently. Thus, biological processes exhibit maximal parallelism. However, since not all computations are explicit parallelizable, the following modes were defined to describe the type of concurrency a computation has: sequential, minimal parallel, bounded parallel, and maximal parallel. In sequential mode, only one rule is used in each computation step. This is because there are computational steps that are inherrently serial because of input-output dependencies. In minimal parallel mode, at least one rule must be used when a set of rules can be used concurrently. In bounded parallel mode, the number of membranes that will compute or the number of rules to be used is restricted. In all modes mentioned,  objects to apply the rule to, as well as the rules themselves, are chosen non-deterministically~\citep{01-paun06,02-paun00,03-gutierrez06,04-martinez09,05-ardelean08,06-wu16}.

In MC, a collection of transitions creates a computation. A computation generates a result as long as it halts, i.e., to have reached to a configuration where no rule can be applied~\citep{01-paun06,02-paun00,03-gutierrez06,04-martinez09,05-ardelean08,06-wu16}.

\section{Comparator P~System}

We now define a membrane that can sort two integers we call a comparator P~System (or~$\Pi_c$ for short). Here, $\Pi_c$~is able to sort two integers~$x$ and~$y$, such that $x = |S_1|$ and $y = |S_2|$. The  multisets~$S_1$ and~$S_2$ are homogeneous multisets, i.e., they contain only one type of objects, and the object in~$S_1$ is different from the object in~$S_2$. For brevity, we represent the multiplicity of the objects in the multisets as $a^x$ to mean that object~$a$ has $x$~copies in the multiset. For example, $s_1 = aaaaa$ and $s_2 = bbb$ respectively represent the integers~5 and~3. The structure of~$\Pi_c$ is $\mu_c = [_{h_0} [_{h_1} ]_{h_1} [_{h_2} ]_{h_2} ]_{h_0}$~\citep{05-ardelean08}.

At the beginning of the computation, two multisets~$a^x$ and~$b^y$ are in~$[_{h_0} ]_{h_0}$. The compartments~$[_{h_1} ]_{h_1}$ and~$[_{h_2} ]_{h_2}$ are both empty. Then, all transactions are performed in two steps in order using the following ruleset:

\begin{enumerate}
\item $R_{1[h_0]}: ab \rightarrow a_{[h_1]}b_{[h_1]}$
\item $R_{2[h_0]}: a  \rightarrow b_{[h_2]}$

      $R_{3[h_0]}: b  \rightarrow b_{[h_2]}$

      $R_{4[h_1]}: b  \rightarrow b_{[h_0]}$
\end{enumerate}

In step~1, the rule~$R_{1[h_0]}$ is applied to the multisets in~$[_{h_0} ]_{h_0}$ translocating an equal number of objects~$a$ and~$b$ to~$[_{h_1} ]_{h_1}$. If~$x>y$, then~$(x-y)$ of~$a$ are left behind at~$[_{h_0} ]_{h_0}$ and all~$y$ of~$b$ are in $[_{h_1} ]_{h_1}$. If~$x=y$, then all objects~$a$ and~$b$ are moved to~$[_{h_1} ]_{h_1}$. If~$x<y$, then $(y-x)$ of~$b$ are left behind at~$[_{h_0} ]_{h_0}$ and all~$x$ of~$a$ are in~$[_{h_1} ]_{h_1}$.

In step~2, the rules~$R_{2[h_0]}$, $R_{3[h_0]}$, and~$R_{4[h_1]}$ can at least be minimally applied in parallel\footnote{At best maximally applied in parallel.}. $R_2[h_0]$~tranlocates the~$a$'s in~$[_{h_0} ]_{h_0}$ to~$[_{h_2} ]_{h_2}$ and at the same time transmutes them as~$b$'s. $R_3[h_0]$~tranlocates the~$b$'s in~$[_{h_0} ]_{h_0}$ to~$[_{h_2} ]_{h_2}$. $R_4[h_1]$~tranlocates the~$b$'s in~$[_{h_1} ]_{h_1}$ to~$[_{h_0} ]_{h_0}$. 

Upon closer investigation, it may seem that~$R_3[h_0]$ has a dependency to~$R_4[h_1]$, in which case we can apply the rules as minimally parallel. However, if we assume that all rules will only fire as long as an object in the corresponding compartment is present, then we can always assume non-dependency and therefore consider the process as maximally parallel. However, when there is only one object~$b$ remaining in~$[_{h_1} ]_{h_1}$, then $R_4[h_1]$ fires first followed by $R_3[h_0]$. This seemingly serial order of the two supposedly concurrent rules instantaneously transfers the remaining~$b$ from~$[_{h_1} ]_{h_1}$ to~$[_{h_2} ]_{h_2}$ through~$[_{h_0} ]_{h_0}$. The time spent, as well as the overhead cost, for passing through transient membrane~$[_{h_0} ]_{h_0}$ are considered zero.

When no more rule among the ruleset can be applied, then the halting state happens. In this case, the larger between $x$ and $y$ will be in $[_{h_2} ]_{h_2}$ while the lower will be in $[_{h_1} ]_{h_1}$~\citep{05-ardelean08}. We call such $\Pi_c$ as a maximizing comparator and is represented as $\Pi_c^+$. 

We can likewise define a minimizing~$\Pi_c$ as~$\Pi_c^-$ where the location of the lower and the higher numbers are reversed in the compartments within~$\Pi_c^-$. In this case, $\max(x,y)$ will be in $[_{h_1} ]_{h_1}$, while $\min(x,y)$ will be in $[_{h_1} ]_{h_1}$.

\section{Membrane Sorter}

We now propose a membrane sorter containing several modified $P_c^+$'s that uses the insertion algorithm to sort a list of $n$~integers $\{x_1, x_2, \dots, x_n\}$, where each integer~$x_i$ is represented as the length~$|S_i|$ of a multiset~$S_i$. We introduce a modification to the~$P_c^+$ described above introducing a rule that implements a cloning out, but with a little twist. We will only allow the clone to copy the first level children compartment of its parent compartment. We define a function $\Root(\mu_1)$ that returns the root of a compartment whose structure is~$\mu_1$. Given that $\mu_1 = [_1 \mu_2 ]_1$, where $\mu_2$ might be flatly- or deeply-rooted structure, then $\Root(\mu_1) = [_1 ]_1$.

Our additional rule which must be triggered conditionally is:
\begin{equation}
  R_{[h_0]}: \Pi_c^+ \rightarrow [_{c_0} [_{c_1} ]_{c_1} \Pi_c^+ ]_{c_0}\label{rule:clone-comparator}
\end{equation}

Here, the compartment~$[_{h_0} ]_{h_0}$ of the parent~$\Pi_c^+$ acts as the~$[_{c_2} ]_{c_2}$ of the cloned~$\Pi_c^+$ such that the structure of the clone~$\mu_c = [_{c_0} [_{c_1} ]_{c_1} [_{h_0} [_{h_1} ]_{h_1} [_{h_2} ]_{h_2} ]_{h_0}  ]_{c_0}$. Note here that the ruleset of the parent is inherited by the clone. Note further that the firing of~$R_{5[h_0]}$ is {\em conditional} such that it will not fire if $[_{h_0} ]_{h_0}$ is already contained in $[_{c_0}]_{c_0}$.

We now present the ruleset that will allow for their recursive and semantically correct implementation:

\begin{enumerate}
\item Follow the ruleset for $\Pi_c^+$ described above.
\item 
  \begin{enumerate}
  \item $R_{5[h_2]}: b \rightarrow b_{[h_0]}$
  \item $R_{6[h_0]}$: Follow the reproduction rule in Equation~\ref{rule:clone-comparator}.
  \item $R_{7[h_0]}: b \rightarrow b_{[c_0]}$
  \end{enumerate}
\item 
  \begin{enumerate}
  \item $R_{8[h_0]}: b \rightarrow b_{[c_0]}$
  \item $R_{9[h_1]}: a \rightarrow b_{[h_0]}$
  \end{enumerate}
\item $R_{10[e]}$ : new multiset $\rightarrow a_{[c_0]}$
\end{enumerate}

Outside of $[_{c_0} ]_{c_0}$ is the environment of the membrane $[_e ]_e$ which contains the multisets whose respective lengths represent the integers to be sorted. The environment member could be $[_{c_0} ]_{c_0}$'s owns cloned outer membrane. When there are no more multisets remaining in $[_e ]_e$ and there are no more rules to implement, then the sorting operation stops. Compartment $[_{h_0} ]_{h_0}$ will have a duality of function: (1)~as the outer membrane of~$\Pi_c^+$ and (2)~as the compartment $[_{c_1} ]_{c_1}$ of the cloned~$\Pi_c^+$. It is important to keep track of the rules to fire so that the meaning will still be semantically correct.

\section{Conclusion}\label{sec:conclude}

This paper presents a membrane-sorter that follows the natural processes of the insertion sort algorithm. The membrane-sorter contains a deeply-rooted comparator P~Systems $\Pi_c^+$'s. Each $\Pi_c^+$ compares the respective sizes of homogeneous multisets and swaps their compartments in levels 1, 2, and 3 of the proposed ruleset. The comparisons and swaps are performed until the input list is sorted. Similar to insertion sort, the membrane-sorter sorts the list online.

\section{Acknowledgment}

The respectfully extend our profound gratitude and appreciation to the adminstrators of Lyceum of the Philippines University -- Laguna (LPU-Laguna), who contributed assistance to the fulfilment of this research paper:
\begin{enumerate}
\item Dr. Peter Laurel -- President
\item Engr. Ricky Bustamante -- Dean, College of Engineering and Computer Studies
\item Mrs. Gerby Muya -- Director, Research and Statistics Center
\end{enumerate}

Without the help and support of these persons, this research might not be performed.

\bibliography{membrane}
\bibliographystyle{plainnat}

\end{document}